\documentclass{aa} 

\def\apjl{ApJ}
\def\apjs{ApJS}
\def\ao{Appl.~Opt.}

\def\aap{A\&A}

\usepackage{graphicx}
\usepackage{txfonts}
\usepackage{graphicx}	
\usepackage{amsmath}	
\usepackage{rotating}
\usepackage{pdflscape}
\usepackage{threeparttable}
\usepackage{color}
\usepackage{hyperref}

\newcommand{\msun}{\,M$_{\sun}$}

\begin{document}

   \title{Recovering chemical bimodalities in observed edge-on stellar disks: insights from AURIGA simulations}

    \author{Francesca Pinna
          \inst{1,}\inst{2,}\inst{3}\thanks{\email{francesca.pinna@iac.es}}
          \and
          Robert J. J. Grand \inst{4,}
          \and 
          Marie Martig \inst{4}
          \and
          Francesca Fragkoudi \inst{5}
          }
   \institute{
   Instituto de Astrofísica de Canarias, Calle Vía Láctea s/n, E-38205 La Laguna, Tenerife, Spain
        \and
Departamento de Astrofísica, Universidad de La Laguna, Av. del Astrofísico Francisco Sánchez s/n, E-38206, La Laguna, Tenerife, Spain
         \and
Max Planck Institute for Astronomy, Koenigstuhl 17, D-69117 Heidelberg, Germany            
\and
Astrophysics Research Institute, Liverpool John Moores University, 146 Brownlow Hill, Liverpool, L3 5RF, UK
\and 
Institute for Computational Cosmology, Department of Physics, Durham University, South Road, Durham DH1 3LE, UK
             }

   \date{Received XXX; accepted XXX}

 
  \abstract
 {
 The well-known bimodal distribution of Milky-Way disk stars in the [$\alpha$/Fe] - metallicity plane is often used to define thick and thin disks.  
 In external edge-on galaxies, there have been attempts to identify this type of bimodality using integral-field spectroscopy (IFS) data. However, for unresolved stellar populations, observations only contain integrated information, making these studies challenging. 
  We assessed the ability to recover chemical bimodalities in IFS observations of edge-on galaxies, using 24 Milky Way-mass galaxies from the AURIGA zoom-in cosmological simulations. 
 We first analyzed the distribution of single stellar particles in the [Mg/Fe] - [Fe/H] plane, finding that bimodality is frequent but not ubiquitous and often unclear. 
 Then, we produced mock IFS [Mg/Fe] and [Fe/H] maps of galaxies seen edge on, and considered integrated stellar-population properties (projected and spatially binned). We investigated how the distribution of stars in the [Mg/Fe] - [Fe/H] plane is affected by edge-on projection and spatial binning. Bimodality is preserved while distributions change their shapes. 
 Naturally, broad distributions of individual star particles are narrowed into smaller [Mg/Fe] and [Fe/H] ranges for spatial bins. 
We observe continuous distributions from high [Mg/Fe] and low [Fe/H], to lower [Mg/Fe] values and higher [Fe/H]. Despite being continuous, these distributions are bimodal in most cases. The overlap in [Fe/H] is small, and different [Mg/Fe] components show up as peaks instead of sequences (even when the latter are present for individual particles).
  The larger the spatial bins, the narrower the [Mg/Fe] - [Fe/H] distribution. 
  This narrowing helps amplify the density of different [Mg/Fe] peaks, often leading to a clearer bimodality in mock IFS observations than for original star particles. 
   We have also assessed the correspondence of chemical bimodalities with the distinction between geometric thick and thin disks. 
   Their individual particles have different distributions but mostly overlap in [Mg/Fe] and [Fe/H]. 
  However, integrated properties of geometric thick and thin disks in mock maps do mostly segregate into different regions of the [Mg/Fe] - [Fe/H] plane. In bimodal distributions, they correspond to the two distinct peaks. Our results show that this approach can be used for bimodality studies in future IFS observations of edge-on external galaxies.}


   \maketitle
%

\section{Introduction}\label{sec:intro}

The chemical bimodality of disk stars in the Milky Way has been a matter of debate for several decades, since two distinct components were found in the [$\alpha$/Fe] - metallicity plane for the solar neighborhood. These 
were initially associated with different disk components defined geometrically (the thick and the thin disk, \citealt{Fuhrmann1998}). 
The chemical bimodality was later used to provide an alternative definition of thick and thin disks, respectively as high- and low-$\alpha$ components \citep[e.g.,][]{Reddy2006,Navarro2011,Adibekyan2012}. 
Large surveys could cover larger volumes of our Galaxy and showed different types of bimodal distributions depending on the analyzed region. 
In the solar neighborhood, they form two $\alpha$ sequences which overlap in metallicity \citep{Hayden2015,Hayden2017}. This overlap is very little at inner radii where the transition between high- and low-$\alpha$ components becomes abrupt  \citep{Haywood2018, Queiroz2020,Queiroz2021}. 

Chemical bimodality has been often interpreted as the result of the formation of the Milky Way disk in different evolutionary stages, as initially proposed in the ''two-infall" model from \citet{Chiappini1997}, and supported by the different corresponding ages \citep[e.g.,][]{Haywood2016, Haywood2018, Hayden2017, Queiroz2023}. This was reaffirmed by 
numerical simulations. 
\citet{Grand2018a} proposed that the high-$\alpha$ component formed in a fast time scale and early on, when gas-rich mergers triggered very intense star-formation phases. The formation of a low-$\alpha$ sequence, on the other hand, would have been initiated by the more metal-poor gas provided by the same important merger. This allowed to reset the metallicity to lower values, and follow a new slower chemical-evolution path, now with a lower abundance of $\alpha$ elements (see also \citealt{Buck2020b} and \citealt{Agertz2021}). 
\citet{Mackereth2018} proposed bursty star formation after intense gas accretion to explain the formation of the high-$\alpha$ sequence, but suggested that chemical bimodality shaped as a double sequence is not very common in Milky-Way mass galaxies. 
Results from \citet{Clarke2019} show that the high-$\alpha$ sequence can form naturally due to intense star formation in clumps, with no need for mergers. If the star formation is spatially more uniform and extended, a low-$\alpha$ sequence is easily formed. 
Thus, there are several proposed scenarios to explain how bimodality arises, each predicting varying degrees of prevalence among spiral galaxies.

High-resolution integral-field spectroscopy (IFS) observations have recently offered the opportunity to study bimodality in external edge-on galaxies. 
\citet{Scott2021} reconstructed the distribution of stars in the [$\alpha$/Fe] - metallicity plane, 
from the weights assigned to different single-stellar-population (SSP) models while fitting the spectrum of each spatial bin with pPXF \citep{Cappellari2004}.  
However, their method faced several challenges compared to resolved stellar-population studies. First, disentangling different stellar populations from integrated light (integrated both in the line of sight and into spatial bins) has large systematic and random uncertainties \citep[e.g.,][]{Sattler2023}. Furthermore, their distribution inherited from the SSP models only two possible values of [$\alpha$/Fe], making the recovery of a full [$\alpha$/Fe] distribution impossible.

In this paper, we propose an alternative method to recover chemical bimodalities in IFS observations: analyzing the distribution of average spatially binned chemical properties. These (mass- or light-) weighted averages have lower uncertainties \citep{Sattler2023} and, although still relying on the use of SSPs, they are less affected by the SSP limited sampling of the [$\alpha$/Fe] space. 
We assess here how results from our method reflect the original (potentially bimodal) distribution of stars, and how easily chemical bimodality can be recovered in IFS observations.
For this purpose, we analyze chemical bimodalities in a sample of 24 simulated galaxies from the AURIGA zoom-in cosmological simulations of Milky Way-mass spiral galaxies. 
This paper is structured as follows. The method is detailed in Sect.~\ref{sec:method}. Our results are gathered in Sect.~\ref{sec:results} and discussed in Section~\ref{sec:discuss}. Section~\ref{sec:conclus} summarizes our conclusions. 

\section{Method}\label{sec:method}

\subsection{AURIGA simulations}
We analyze here a sample of 24 simulated galaxies selected by \citet{Pinna2023} from the AURIGA zoom-in cosmological simulations \citep{Grand2017,Grand2024}. 
These simulations were performed with an improved version of the moving-mesh AREPO code \citep{Springel2010, Pakmor2016,Weinberger2020}, which includes magneto-hydrodynamics and collisionless dynamics. 
From the entire suite of simulations, we used resolution at ''level 4'', providing the largest sample of 30 galaxies. 
This level ensures a mass resolution of $\sim 3 \times 10^5$\msun\,and $\sim 5 \times 10^4$\msun, respectively for dark-matter and baryonic-mass particles. 
These zoom-in simulations result from re-simulating 30 haloes, selected from the parent 100-Mpc cubed volume Dark Matter Only EAGLE simulation \citep{Schaye2015}, to have a mass between $1$ and $2 \times 10^{12}$ \msun and to be relatively isolated at redshift $z=0$. 
Once a halo is identified, the zoom-in initial conditions are created at redshift $z=127$ using public Gaussian white noise field realization Panphasia \citep{Jenkins2013}. This creates a refined particle distribution within the Lagrangian region of the selected halo of the order of a Mpc in size, which is surrounded by a lower-resolution particle distribution at larger distances. Each dark matter particle is then split into a dark matter particle-gas cell pair, whose mass and center of mass are determined by the cosmic baryon fraction. 
 
The interstellar medium is represented by a subgrid model with two phases (cold and dense clouds in a hot medium, \citealt{Springel2003}). Star formation happens when the gas reaches the density threshold of 0.13~cm$^{-3}$ and becomes thermally unstable. In that case, the gas cell is either converted into a star particle (a single stellar population with specific age and chemical abundances), or becomes the site for a supernova type II (SNII) feedback, following a distribution of stellar masses given by a \citet{Chabrier2003} initial mass function. 
 The total metals associated with a SNII are given by the initial metallicity of the gas cell plus the new metals, produced following yield prescriptions from \citet{Portinari1998}. 
In a SNII site, a wind particle is launched, loaded with $(1 - \eta)$ of the total metals, where the metal loading factor is $\eta = 0.6$. The rest of the total metals remain in the original gas cell site. 
The mass loss and metal content from supernovae type Ia and asymptotic giant branch (AGB) stars were also calculated, using yield tables from \citet{Thielemann2003}, \citet{Travaglio2004} and \citet{Karakas2010}, and distributed among nearby gas cells. Nine specific chemical elements were tracked during the simulations: H, He, C, O, N, Ne, Mg, Si, and
Fe. 
The chemical enrichment model used for AURIGA simulations was shown to be successful in several studies \citep[e.g.,][]{Grand2018a, Grand2019, Grand2020, vandeVoort2020}. 

Black-hole seeds of $10^5$\msun $h^{-1}$ were placed at the center of haloes once they became more massive than $ 5 \times 10^{10} {\rm M_{\sun}} h^{-1}$, at the position of the densest gas cell. These seeds acquire mass by accreting nearby gas cells or merging with other black holes. Nuclear activity via accreting black holes is modeled via the injection of thermal energy to the surrounding gas cells in two separate modes: a high-accretion rate quasar mode and a ''maintenance'' radio mode \citep[see also][for a study on the effects of these modes on galaxy properties]{Irodotou2022}. 
For further details about the simulations, we refer the reader to \citet{Grand2017}. 

These simulations provide properties of individual star particles (including single-element chemical abundances) at each snapshot. Here we use such properties, in particular metallicities and [Mg/Fe] abundances, at redshift $z=0$, at the end of the simulations. 
In this paper, we use [Mg/Fe] as a proxy for [$\alpha$/Fe]. 
Ages of star particles are also available. While a detailed analysis of age distributions is beyond the scope of this paper, we use stellar ages to extract star-formation histories, providing another missing piece of the puzzle. We show and discuss two examples in Sect.~\ref{sec:discuss}, while a detailed study of star-formation histories is deferred to a companion paper.

\subsection{Galaxy sample, projection, binning and spatial selection} \label{sub:defin}

Our sample includes the 24 galaxies selected by \citet{Pinna2023} from the initial AURIGA sample of 30 galaxies. Six galaxies with strong distortions, ongoing major mergers, or without a clear disky shape were excluded. 
The 24 galaxies are spirals with morphological types from Sa to Sc \citep{Walo2021}, of which 19 are barred, and masses between $10^{10}$\msun\,and $10^{11}$\msun. They have different sizes, with optical radii $R_{opt}$ between 13 and 37~kpc \citep{Grand2017}. 
The main goal of this paper is to analyze bimodalities from projected and spatially binned chemical properties, which we would have in stellar-population maps from IFS observations. 
For this purpose, following \citet{Pinna2023}, we projected to an edge-on view the 24 galaxies in our sample, integrating along the line of sight the properties of all star particles within an intrinsic radius of 40~kpc, into pixels of 200-pc size. Then, we selected only pixels with particles, and applied a Voronoi binning \citep{Cappellari2003} to ensure a minimum number of particles per bin, which depended on the binning level. 
We used three different binning levels, with a target of 16, 36 and 900 particles per Voronoi bin, leading to respective average bin sizes of 0.3~kpc, 0.7 to 0.9~kpc, and 1.8 to 2.1~kpc (depending on the galaxy). 
These are typical bin sizes in IFS observations of edge-on galaxies \citep{Comeron2019,Pinna2019b,Pinna2019a,Martig2021,Sattler2023,vandeSande2023}. 

We included Voronoi bins in a region within a distance of $\pm 2h_{scale}$ from the midplane and a projected radius of $0.8 R_{opt}$. 
This follows \citet{Pinna2023} who analyzed the same galaxy sample in a projected region of the same height and radius. 
$h_{scale}$ is the standard deviation of the vertical positions of stellar particles at $0.5 R_{opt}$ \citep{GarciadelaCruz2021}, and ranges from 2.1 to 13.1~kpc \citep{Pinna2023}. $R_{opt}$ is the optical radius (the radius where the $B$-band surface brightness drops below 25~mag arcsec$^{-2}$), and ranges from 13.5 to 37~kpc \citep{Grand2017}. 
Bins located in regions where the disk does not dominate were not considered (following \citealt{Pinna2023}): within $R_{0,disk}$, the radius where the disk starts to dominate over the central component, and closer than $z_{tT}$ to the midplane, where $z_{tT}$ is the transition height between the thick and the thin disk (see Sect.~\ref{sub:defthickthin}). 
[Mg/Fe] and [Fe/H] were calculated as a mass-weighted average of particle properties in each bin. 
For additional details about the galaxy sample, the projection and binning process, we refer the reader to \citet{Pinna2023}. 

We analyze in Sect.~\ref{sec:results} how average projected and Voronoi binned chemical properties reflect the properties of the original star particles of the simulations. 
To do this, we compare the distribution in the [Mg/Fe] - [Fe/H] plane of Voronoi bins with the original star-particle distribution. 
Star particles were selected for this purpose in a region as similar as possible to Voronoi bins, with the same spatial limits but in the three spatial dimensions, using the intrinsic radius instead of the projected radius. 
Thus, we included particles inside a cylinder of height $4h_{scale}$ (with the galaxy midplane located at $2h_{scale}$), and intrinsic radius $0.8 R_{opt}$. 
Following \citet{Pinna2023} and our selection for Voronoi bins, particles contained within a radius of $R_{0,disk}$, at thin-disk heights (see Sect.~\ref{sub:defthickthin}), 
were not taken into account. 

\subsection{Definition of thick and thin disks} \label{sub:defthickthin}
In this paper, we use the geometric definition of thick and thin disks in this same sample of galaxies from \citet{Pinna2023}. They performed a morphological decomposition by fitting vertical surface brightness profiles with a double hyperbolic secant square, representing the two disk components. 
All profiles except the ones of the galaxy Au1 were fitted well with these two components. 
For Au1, we used only one component and concluded that this galaxy does not have a clear distinction between thick and thin disks. 
This decomposition allows to define a region dominated by the thin disk, close to the midplane, and a region dominated by the thick disk, at larger heights. The transition between them happens at $z_{tT}$, where the two hyperbolic secant squares cross each other. The values for $z_{tT}$ are indicated in Table~A.1 in \citet{Pinna2023}. 
\section{Results}\label{sec:results}

\begin{figure*}
\centering
\resizebox{1.\textwidth}{!}
{\includegraphics[scale=1.]
{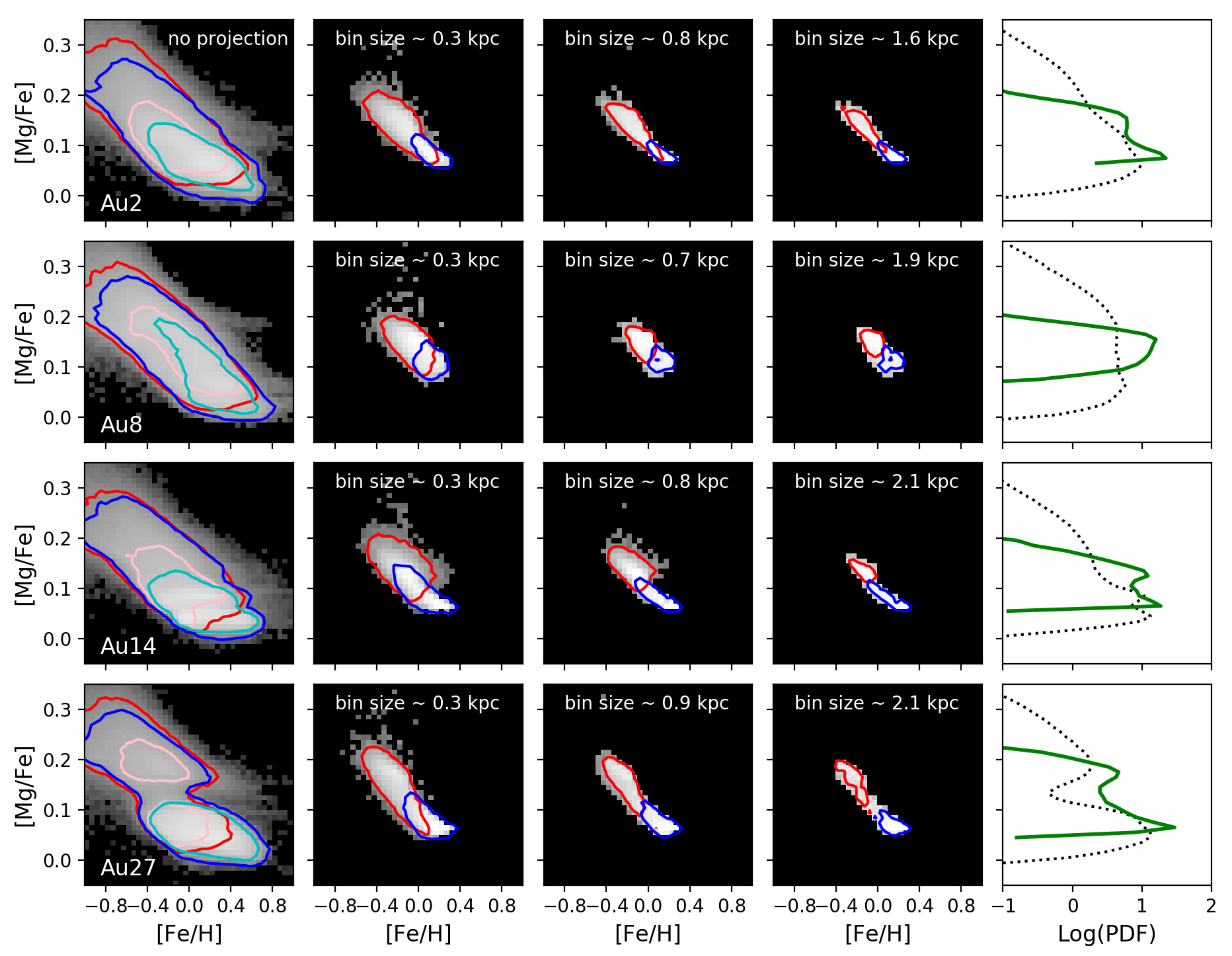}}
\caption{Mass-weighted distribution (in terms of probability distribution function, PDF) of disk stars in the [Mg/Fe] - [Fe/H] plane, for four galaxies in our sample. Each row corresponds to one galaxy, whose name is indicated on bottom-left of the leftmost panels. 
From left to right, {\it first column}: properties of individual star particles. 
{\it Second to fourth column}: stellar properties in mock edge-on IFS maps, using different Voronoi-binning levels. 
Average bin sizes range from 0.3 to 2.1 kpc (indicated on top of each panel). 
{\it Fifth column}: one-dimensional PDFs of [Mg/Fe] abundances for individual star particles in the leftmost panels (black dotted lines), and for Voronoi bins in the second column from left  (green solid lines).
All PDFs are shown in logarithmic scale. Red and blue (pink and cyan) contours in the two-dimensional distributions enclose the 50\% (16\%) highest-density points, respectively in the thick and the thin disks. 
The remaining 20 galaxies of the sample are displayed in Appendix~\ref{app:bimod_sample}. 
}
\label{fig:bimod_4gal}
\end{figure*}

In this section, we describe how the distribution in the [Mg/Fe] - [Fe/H] plane changes from using individual star particles to using integrated properties in spatial bins, which are what is available in IFS observations of unresolved stellar populations. 
We show in Fig.~\ref{fig:bimod_4gal} four galaxies that are representative of the variability of the distribution shapes, 
while the remaining galaxies of the sample are shown in Appendix~\ref{app:bimod_sample}. 
These four galaxies represent four groups in which the full sample was classified based on the shape of their star-particle and Voronoi-bin distributions (see Sect.~\ref{sub:class}). 
In the leftmost panels of Fig.~\ref{fig:bimod_4gal} we plotted, for comparison, probability density functions (PDFs) of the properties of individual stellar particles in the disk (for the selected volume and disk definition, see Sect.~\ref{sub:defin}). 
The one-dimensional PDFs of the [Mg/Fe] abundance of individual particles are represented in the rightmost panels as black dotted lines. 
In the second to fourth columns from left to right, we show the PDFs of the chemical properties of Voronoi bins for different binning levels (increasing bin size from left to right, Sect.~\ref{sub:defin}). 
The one-dimensional [Mg/Fe] PDFs corresponding to the lowest level of binning (second column from left) are indicated with green solid lines in the rightmost panels of Fig.~\ref{fig:bimod_4gal}. 
All PDFs were normalized such that the integral over the plotted range is 1, and shown in logarithmic scales. 

\subsection{Distribution of individual star particles in the [Mg/Fe] - [Fe/H] plane} \label{app:bimod_allpart}

We focus here on two-dimensional distributions (with the two dimensions being [Fe/H] and [Mg/Fe]) of individual stellar particles in the [Mg/Fe] - [Fe/H] plane (leftmost panels of Fig.~\ref{fig:bimod_4gal}), and one-dimensional [Mg/Fe] distributions in the rightmost panels (black dotted line). 
In two galaxies in Fig.~\ref{fig:bimod_4gal}, Au2 and Au8, individual star particles show smooth distributions (two top rows). This does not mean that higher- and lower-[Mg/Fe] components are not present. The distribution in Au2 (black dotted line in the top rightmost panel) is the composition of a main low-$\alpha$ component with two higher-$\alpha$ humps. The broad Au8 distribution might also be the composition of two contributions of different [Mg/Fe] and similar density, although this is not clear. 
Bimodality or multimodality is sharply clear in the other two galaxies. Au14 has two low-$\alpha$ sequences of similar density (below [Mg/Fe]$\sim 0.1$ dex), overlapping in [Fe/H], and a third component at higher $\alpha$ (with a hump shape). The latter is not fully separated and covers a wide [Mg/Fe] and [Fe/H] range. 
Finally, Au27 shows two very distinct thick sequences, with different [Mg/Fe] ranges and mostly overlapping in metallicity. This corresponds to a very clear bimodal [Mg/Fe] distribution in the bottom, rightmost panel of Fig.~\ref{fig:bimod_4gal} (black dotted line). 
Results in the leftmost panels of Fig.~\ref{fig:bimod_4gal} (but also in Fig.~\ref{fig:bimod_sample0}, \ref{fig:bimod_sample1} and \ref{fig:bimod_sample2}) are in perfect agreement with distributions of star particles by \citet{Grand2018a}. It is important to note that they used AURIGA simulations at resolution ''level 3'' (higher resolution than we used here), and they sliced stellar particles in different radial and vertical bins. We used here a different approach analyzing the full galaxy including different radii and heights, and overlaying contours based on geometrical thick- and thin-disk definitions. 
In particular, \citet{Verma2021} had already found a clear chemical dichotomy in the simulation at ''level 3'' of Au27, especially if compared with Milky-Way observed bimodality. 

\subsection{Chemical bimodality in mock integral-field spectroscopy (IFS) observations}\label{sub:bimod_binned}

In general, the four galaxies in Fig.~\ref{fig:bimod_4gal} display continuous two-dimensional distributions with slightly different shapes (first four columns on the left). 
Average values of [Mg/Fe] and [Fe/H] for Voronoi bins follow continuous (but mostly bimodal) distributions from higher [Mg/Fe] and lower [Fe/H] to lower [Mg/Fe] and higher [Fe/H]. Compared to the distribution of individual star particles (leftmost column), 
Voronoi-bin distributions naturally cover smaller [Mg/Fe] and [Fe/H] ranges. In particular, $\alpha$ sequences that cover a wide overlapping [Fe/H] range in the distribution of individual particles turn into a bimodal but continuous and narrow distribution of Voronoi bins (Au14 and Au27).  
No clear $\alpha$ sequences overlapping in metallicity are observed for Voronoi bins, suggesting that this kind of bimodality will be hardly identified in observations of unresolved stellar populations. 
However, the one-dimensional [Mg/Fe] PDF is clearly bimodal in three of the galaxies (all except Au8, green solid line). 
Galaxies with clearer bimodality in the particle distribution, show more segregated [Mg/Fe] peaks in the bimodal distribution of Voronoi bins (see e.g. comparison between the Au2 and Au27). 
The use of integrated properties can actually facilitate the identification of bimodalities. 

Distributions of components with different $\alpha$ become narrower (in both [Fe/H] and [Mg/Fe]) for Voronoi bins than for individual particles. For example, in Au2 [Mg/Fe] double peaks appear clearer for Voronoi binning even though different $\alpha$ components did not appear clearly separated for star particles. 
The double low-$\alpha$ sequence in Au14 turns into one low-$\alpha$ peak, while a second peak located at higher $\alpha$ (green solid line) includes star particles in the highest-$\alpha$ hump (black dotted line). 
On the other hand, the broad distribution of individual star particles in Au8 only turns into a narrower unimodal distribution. 

Figure~\ref{fig:bimod_4gal} also shows that the tracks in [Mg/Fe] become narrower for larger Voronoi bins, (third and fourth columns from the left). 
However we have checked that, while they become narrower, peaks keep the same central [Mg/Fe]. 

\subsection{Galaxy classification} \label{sub:class}

Galaxies in our sample were classified into four groups, based on the comparison of Fig.~\ref{fig:bimod_sample0}, \ref{fig:bimod_sample1} and \ref{fig:bimod_sample2} with Fig.~\ref{fig:bimod_4gal}. 
 The four groups are:

\begin{enumerate}
    \item {\it Group 1:} Galaxies that show a clear bimodal distribution both for star particles and spatial bins in mock IFS maps. Represented by Au27 in Fig.~\ref{fig:bimod_4gal}, it includes Au5, Au7, Au9, Au12, Au24.  
    \item {\it Group 2:} Galaxies with two peaks plus one (or more) additional components in the particle distribution and a bimodal distribution in mock maps. This group is represented by Au14 in Fig.~\ref{fig:bimod_4gal}, and includes Au10, Au17, Au18, Au22, Au26. 
    \item {\it Group 3:} Galaxies with clear bimodality in mock IFS observations, when this bimodality in the particle distribution is not sharply clear, although black-dotted lines suggest they are the combination of two or more components. This group is represented by Au2 in Fig.~\ref{fig:bimod_4gal}, and includes Au3, Au6, Au15, Au19, Au21, Au23, Au25. 
    \item {\it Group 4:} Galaxies without a clear bimodality in the particle distribution nor in mock IFS maps. This group is represented by Au8 in Fig.~\ref{fig:bimod_4gal}, and includes Au16 and Au28.  
\end{enumerate}

\subsection{The connection between chemical bimodality and the distinction between thick and thin disks} \label{sub:thickthin}

In this section, we analyze the connection of different components in the [Mg/Fe] - [Fe/H] plane with thick and thin disks (see Sect.~\ref{sub:defthickthin} for their definition). We have overlapped in Fig.~\ref{fig:bimod_4gal} red and blue contours of the 50\% percentiles of the points (of the individual particles in the leftmost column and of the Voronoi bins in the second to fourth columns from the left), respectively in the thick and thin disks. 
Red and blue contours show that the star-particle distributions of thick- and thin-disk stars in the [Mg/Fe] - [Fe/H] plane mostly overlap. However, 84\% percentiles of the particle distributions (enclosing the highest-density 16\% of the particles) for thick (pink contours) and thin disks (in cyan), show that they peak in different regions, although they do not segregate in clearly different $\alpha$ components.
Thick-disk stars mostly contribute to a higher-$\alpha$ region, while thin-disk stars tend to contribute more to the low-$\alpha$ components. 
In Au27, the thin-disk distribution is mostly concentrated in the lower-$\alpha$ sequence (cyan contour). Nevertheless, the (geometric) thick disk gives a strong contribution to both sequences (pink contours), probably because of the warp of the low-$\alpha$ component (Fig.~5 in \citealt{Pinna2023}). 
The thin disk in Au14 covers the two low-$\alpha$ sequences with the 84\%-percentile contour and contributes less to the high-$\alpha$ region. 
We also note that the overlapping between thin and thick disks is not necessarily larger in galaxies with no clearly distinct components in the [Mg/Fe] - [Fe/H] plane (e.g., Au8).

Figure~\ref{fig:bimod_4gal} shows that the chemical separation between the geometric thick and thin disks improves when using integrated properties. In bimodal distributions (Au2, Au14 and Au27), there is a clear correspondence between thick and thin disks and the two peaks, respectively the high-$\alpha$ and the low-$\alpha$ components. 
In Au8, instead, thick and thin disks cover different regions of the same unimodal distribution (respectively at high [Mg/Fe] and low [Fe/H], and at low [Mg/Fe] and high [Fe/H]). 
In general, the larger the bin size, the lower the overlapping between the thin and thick disks. We show here that projection and binning facilitate the separation of geometric thick- and thin-disk distributions in the [Mg/Fe] - [Fe/H] plane. 

\section{Discussion}\label{sec:discuss}

\subsection{Chemical bimodalities and galaxy mass assembly} \label{sub:mass_assembly}

Our results show that distributions of stars in the [Mg/Fe] - [Fe/H] plane are continuous (but this does not imply unimodal), and thick and thin disks correspond to the two high- and low-$\alpha$ ends of the same continuous distributions. 
This indicates that star formation did not necessarily completely quench globally in the disk at any time (see star-formation histories and discussion below), nor at the transition between the thick-disk and the thin-disk formation. Alternatively, accretion may have provided stars of intermediate chemical abundances between stars formed in different starbursts, and blurred the separation between stars formed in these bursts both in ages and chemical properties. 
On the other hand, different peaks, blobs and sequences in the [Mg/Fe] - [Fe/H] plane are found in most of these continuous distributions and point to different intense star-forming episodes. 
These distinct high-density components in the [Mg/Fe] - [Fe/H] plane are naturally interpreted as populations of stars formed in different phases, potentially with different efficiencies and time scales, from gas with different chemical abundances (e.g., \citealt{Hayden2015}, \citealt{Grand2018a}, \citealt{Queiroz2023}, see also Sect.~\ref{sec:intro}). 
These different star-forming episodes did not involve necessarily only a specific vertical range, or one disk morphological component, but globally the galaxy at all heights. Nevertheless, they had a
different impact, in most galaxies, at specific heights (thus, either on the thick or the thin disk). This is 
indicated by the overlapping of thick- and thin-disk particle distributions in the [Mg/Fe] - [Fe/H] plane. 
Furthermore, some thin or thick disks are themselves made up of different $\alpha$ components (Au10, Au14), suggesting that each one may have been assembled by multiple star-formation bursts. 
Our results are in agreement with previous numerical studies showing that thick and thin disks defined as two different chemical components do not necessarily correspond strictly to the geometrically defined thick and thin disks \citep[e.g.,][]{Agertz2021}. 

The presence of multiple components in the [Mg/Fe] - [Fe/H] plane was previously connected to violent events such as strong starbursts produced by gas-rich mergers \citep[e.g.,][]{Grand2018a, Grand2020}. Interestingly, the galaxy with the strongest bimodality in Fig.~\ref{fig:bimod_4gal}, Au27, had a relatively quiescent merger history. The last important merger happened around 9~Gyr ago and may have led to the formation of the high-$\alpha$ sequence, with no other subsequent events \citep{Grand2018a}. 
Among galaxies in Fig.~\ref{fig:bimod_4gal}, it is the one with the lowest fraction of ex-situ stars in the analyzed region ($\sim 8.5$\%, \citealt{Pinna2023}). It was classified in a subsample of ``very small or negligible fraction of ex-situ material'' by \citet{Gomez2017a}. 
On the other hand, the only galaxy in Fig.~\ref{fig:bimod_4gal} not showing bimodality, Au8, is the one with the highest accreted mass fraction among the four ($\sim 15.5$\%), according to \citet{Pinna2023}. 
Repeated mergers might lead to a mixing of different (in-situ and ex-situ) populations of stars with different chemical abundances, diluting bimodalities. Repeated gas-rich mergers may also fuel continuous star formation without necessarily resetting chemical properties. 
This was the case of Au8, with at least nine important mergers \citep{Monachesi2019}. 

This does not mean that a strong merger contribution leads necessarily to a unimodal distribution (see e.g. Au7, 
Appendix~\ref{app:bimod_sample}). 
Au7 shows a clear bimodality (Group 1,  Sect.~\ref{sub:class}) and is one of the galaxies with the highest accreted fraction in our full sample ($\sim 29$\%, \citealt{Pinna2023}). This example shows that active merger histories can also lead to strong bimodalities. The most significant satellite contributor to the ex-situ component in Au7 was almost twice more massive than the one in Au8 \citep{Gomez2017a}, also with an active merger history but in Group 4. The most important merger happened in Au7 about 2~Gyr later than in Au8, and with a highly inclined orbit. In Au24, also in Group 1, the last significant merger happened with a similar inclination as Au7, although at a similar time as Au8 but accreting a much lower mass. 

\begin{figure*}
\centering
\resizebox{1.\textwidth}{!}
{\includegraphics[scale=1.]
{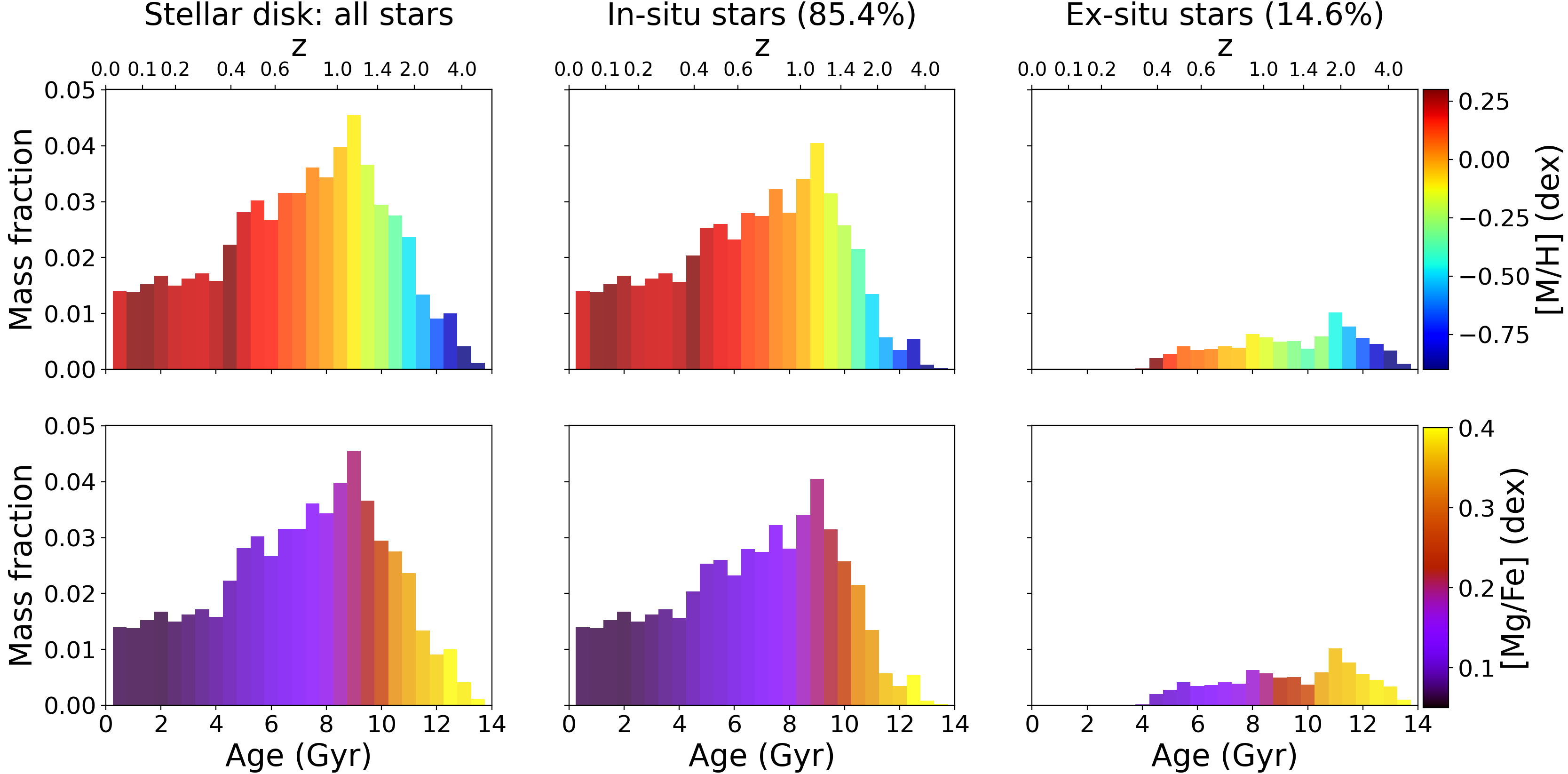}}
\caption{Star-formation history in terms of mass fraction per age bin, of the stellar disk (Sect.~\ref{sub:defthickthin}) in the galaxy Au8. Age bins (of a half-Gyr width) are color coded by the average [Fe/H] metallicity in the top panels and by the average [Mg/Fe] abundance in the bottom panels. The leftmost panels show the mass fraction in each age bin for the entire stellar disk. This mass fraction is split into the in-situ (middle panels) and the ex-situ stars (rightmost panels). Age can be interpreted as the look-back time when the corresponding mass fraction was formed. The corresponding redshift is indicated in the top axes. 
All star-formation histories were normalized to the galaxy mass. Total mass fractions of the in-situ and ex-situ components are indicated on top of each panel. 
}
\label{fig:sfh_au8}
\end{figure*}

\begin{figure*}
\centering
\resizebox{1.\textwidth}{!}
{\includegraphics[scale=1.]
{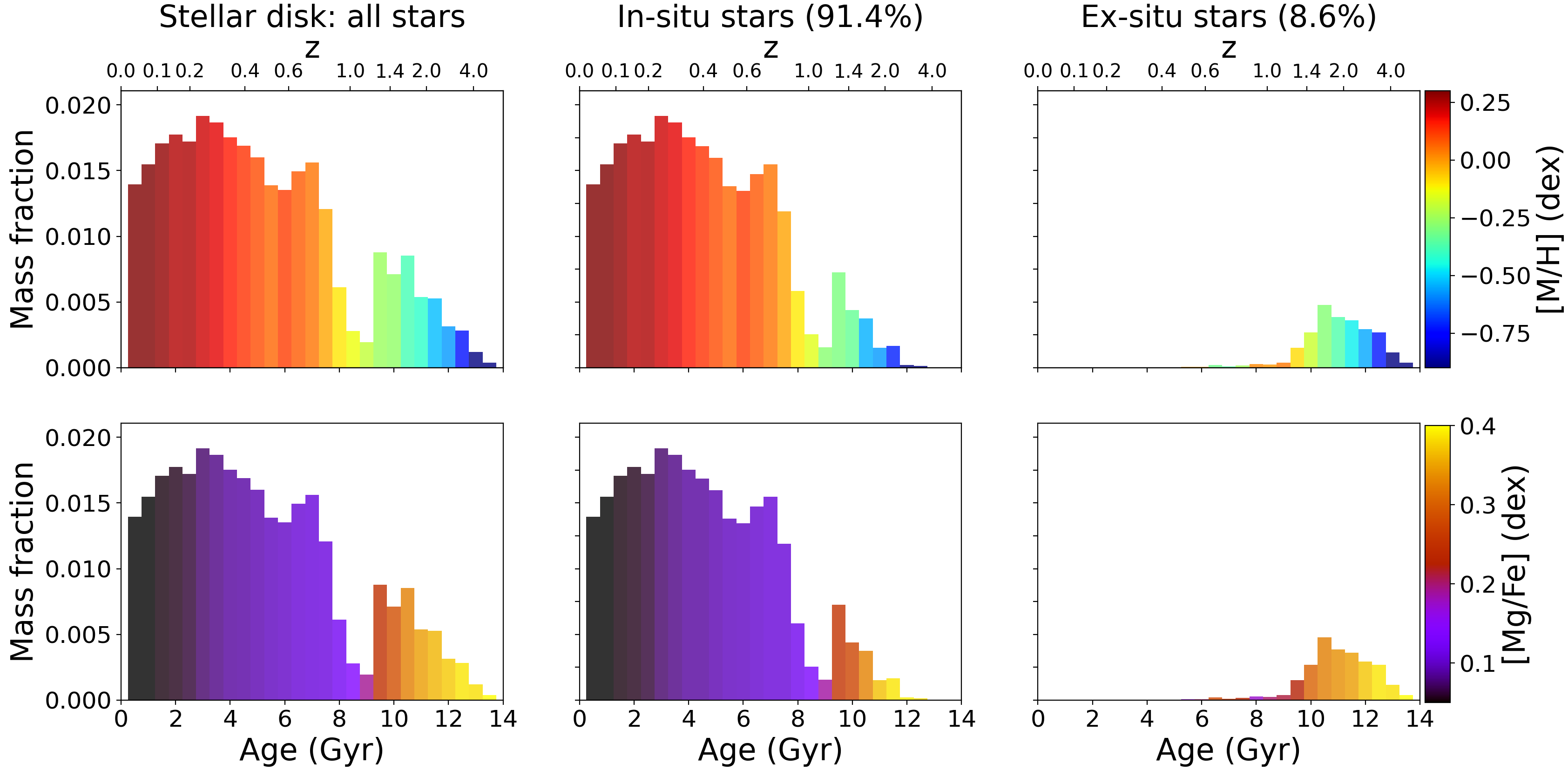}}
\caption{Same as Fig.~\ref{fig:sfh_au8} but for the galaxy Au27. 
}
\label{fig:sfh_au27}
\end{figure*}

This discussion is better illustrated by star-formation histories, which connect properties at redshift $z=0$ in Fig.~\ref{fig:bimod_4gal} with mass assembly at different times. We show two key representative examples in Fig.~\ref{fig:sfh_au8} and \ref{fig:sfh_au27}, relative respectively to Au8, the galaxy in Fig.~\ref{fig:bimod_4gal} with a unimodal distribution, and Au27, the one with the clearest bimodality in Fig.~\ref{fig:bimod_4gal}. Star-formation histories of the full sample of galaxies will be presented in a companion paper. 
In Fig.~\ref{fig:sfh_au8} and \ref{fig:sfh_au27}, star-formation histories are plotted in terms of the mass fraction of stars of a specific age in their stellar disks defined as in Sect.~\ref{sub:defin}. Age bins are color coded by [Fe/H] in the top panels and [Mg/Fe] in the bottom panels. The entire stellar disk is shown in the leftmost panels. Au8 shows a relatively continuous star-formation history, peaking around 9~Gyr ago, corresponding to a smooth and continuous chemical evolution. Au27 shows two distinct strong peaks with a drop in the mass fraction at ages of about 8.5~Gyr. 

It is necessary to split the mass fraction into in-situ and ex-situ components (respectively middle and rightmost panels), to understand to what extent this behavior is driven by internal star formation or star accretion. Au8 does not show clearly distinct contributions in the star-formation history either from the in-situ and ex-situ stars. Both components show a rather continuous chemical evolution. 
In Au27, the in-situ component is the main driver of the disk bimodality, with two peaks of internal star formation with clearly different chemistry. The oldest is metal poor and $\alpha$ enhanced, while the youngest is metal rich and has much lower [Mg/Fe]. The ex-situ component is dominated by a small satellite that was accreted at relatively early times (around 8~Gyr ago) and contributed to the first strong peak of metal-poor and $\alpha$-enhanced stars. Just after the merger, the most intense burst of internal star formation happened, transitioning to higher metallicity and lower [Mg/Fe]. This merger very likely provided a large amount of gas from the satellite galaxy to fuel the second starburst. 
In general, galaxies with clear bimodalities had rather bursty (in-situ) star-formation histories. Au14 had three massive peaks of star formation. These were probably responsible for the formation of the three components in their [Mg/Fe] - [Fe/H] star-particle distributions.  The later important star-formation event corresponded to the extended formation of the low-$\alpha$ component. In Au14, the three starbursts may correspond to the pericentric passage of a satellite (see also Fig.~B.19 in \citealt{Pinna2023}). These results are in total agreement with previous work by the AURIGA team \citep{Grand2018a,Grand2020}. 

\subsection{Observing chemical bimodalities in edge-on stellar disks} \label{sub:disc_obs}

This discussion highlights the importance of identifying potential chemical bimodalities of thick and thin disks in external galaxies. 
Our results have shown here that, while double-sequence structures are unlikely to be observed in unresolved stellar populations, the edge-on projection actually facilitates the identification of bimodal, double-peak, [$\alpha$/Fe] distributions. 
Bimodal distributions in IFS observations result from the presence of different components, partially overlapping and not necessarily clearly distinct, in the distribution of individual stars whose light is integrated into the Voronoi bins. 
Bimodalities identified observationally with our method (using integrated properties) usually correspond to the geometric thick and thin disks, which generally formed in two subsequent phases of the evolution of the galaxy. 
On the other hand, the lack of a bimodal [Mg/Fe] distribution in some galaxies suggests that their thick and thin disks result from a continuous and smoother formation. 

A two-peak distribution was also found by \citet{Scott2021}, in IFS observations of an edge-on spiral galaxy, with a different method. They used the [$\alpha$/Fe] - metallicity distribution recovered from the weights assigned to different SSP models during the spectral fitting (see also Sect.~\ref{sec:intro}), for each one of their large spatial bins located at different radii and distances from the midplane. 
We propose here, instead, to use the distribution of all average [$\alpha$/Fe] and metallicity values from smaller spatial bins (each value extracted from one spectrum), and finally analyze the correspondence of peaks in bimodal distributions with thick and thin disks. 
Our method can mitigate the impact of large uncertainties of stellar-population parameters in their distributions extracted from the same spectrum (see e.g. Fig.~B3 in \citealt{Sattler2023}), as well as the lack of good [Mg/Fe] coverage of some SSP models \citep[e.g.,][]{Vazdekis2015} by using (weighted) average [Mg/Fe] values of each spatial bin (e.g., values in Fig.~3 in \citealt{Pinna2019a}). 


\section{Conclusions}\label{sec:conclus}

In this paper, we have investigated how easily bimodalities in the [Mg/Fe] - [Fe/H] plane can be recovered from projected and spatially binned information that are available in IFS observations, and to what extent this bimodality matches the geometrically defined thick and thin disks. 
From our results, we draw the following conclusions:
\begin{itemize}
     \item Chemical bimodality is not ubiquitous in galaxies. In some galaxies, stars are distributed in two or more sequences in the [Mg/Fe] - [Fe/H] plane, overlapping in [Fe/H]. In other galaxies, different components in the [Mg/Fe] - [Fe/H] plane are not clearly distinct and/or do not form sequences. Finally, some other galaxies do not show distinct [Mg/Fe] - [Fe/H] components. 
     \item In general, bimodality is enhanced in projected and spatially binned information of mock IFS maps.  In most galaxies, the presence of two or more [Mg/Fe] components in the star-particle distribution before projection, even when these are not clearly separated, translates into a double-peaked distribution of Voronoi bins. 
   \item Star particles belonging to the geometrically defined thick and thin disks are not traced by different sequences in the [Mg/Fe] - [Fe/H] plane, and thick- and thin-disk distributions mostly overlap. However, the two peaks in the [Mg/Fe] - [Fe/H] distribution of projected Voronoi bins correspond to the distributions of the geometric thick and thin disks (albeit still with some overlapping between them). 
   \item A more aggressive spatial binning (larger bins) leads to narrower distributions with thick and thin disks that are more chemically segregated. 
\end{itemize}
We propose to use the distribution of average [Mg/Fe] and [Fe/H], calculated for each Voroni bin, to investigate bimodalities in observations of edge-on external galaxies. We have shown that this method facilitates the recovery of bimodalities and helps to extract important information about the different formation phases of thick and thin disks. 

\section*{Acknowledgements}
FP acknowledges support from the Spanish Ministry of Science and Innovation (MICINN) through the project PID2021-128131NB-I00/10.13039/501100011033. FP would like to thank Anna Queiroz for the useful discussion. 
RG acknowledges financial support from an STFC Ernest Rutherford Fellowship (ST/W003643/1). 
FF acknowledges support from a UKRI Future Leaders Fellowship (grant no. MR/X033740/1).

\bibliographystyle{aa}
\bibliography{biblio_bimodality.bib}

\begin{appendix} 

\section{Chemical bimodalities from integrated properties for our full sample} \label{app:bimod_sample}

\begin{figure*}
\centering
\resizebox{1.\textwidth}{!}
{\includegraphics[scale=1.]{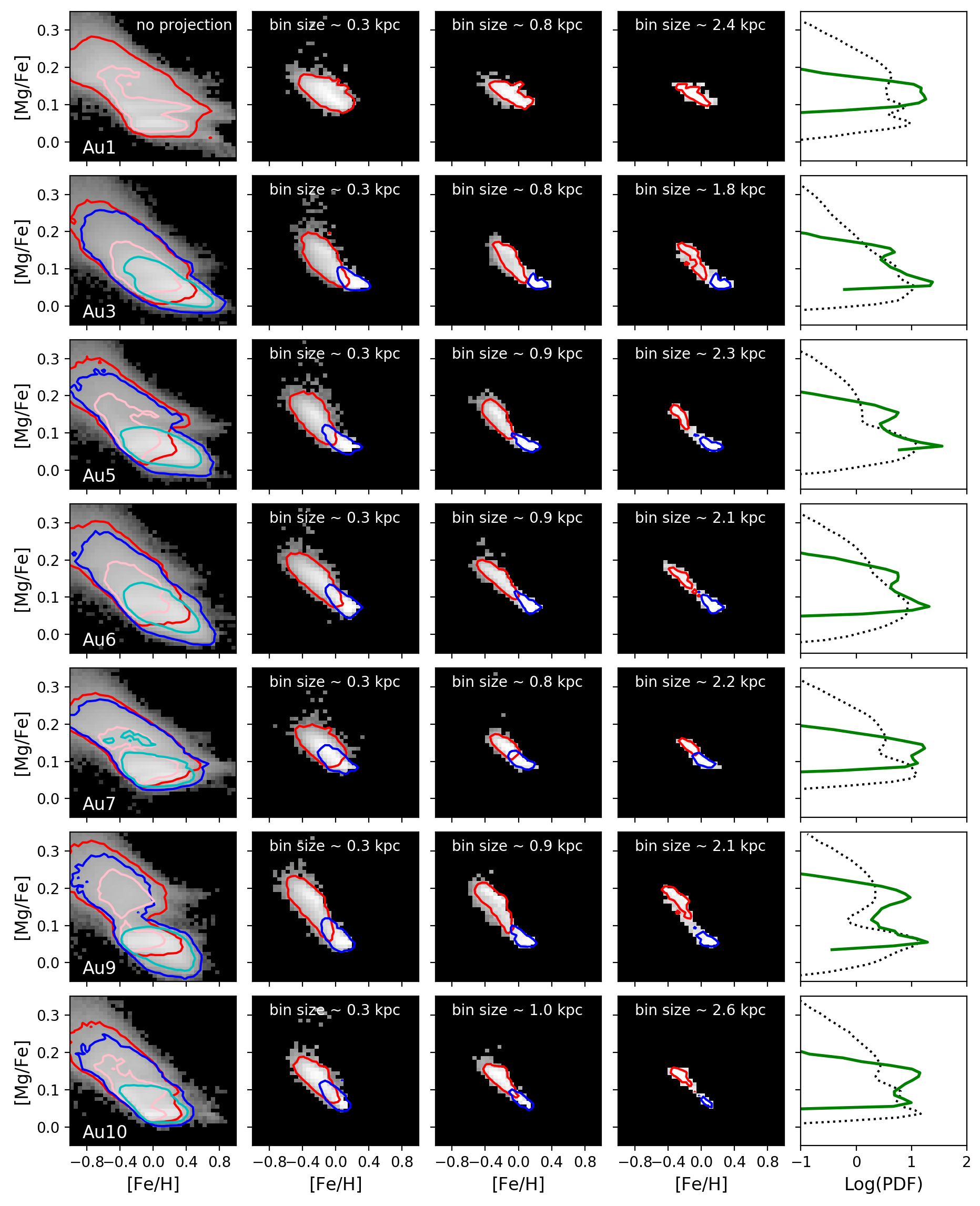}}
\caption{Same as Fig.~\ref{fig:bimod_4gal} for galaxies Au1, Au3, Au5 to Au7, Au9 and Au10. Au1 does not have a clear double-disk structure and red (pink) contours indicate the 50\% (84\%) percentiles of the full disk distribution. 
}
\label{fig:bimod_sample0}
\end{figure*}

\begin{figure*}
\centering
\resizebox{1.\textwidth}{!}
{\includegraphics[scale=1.]{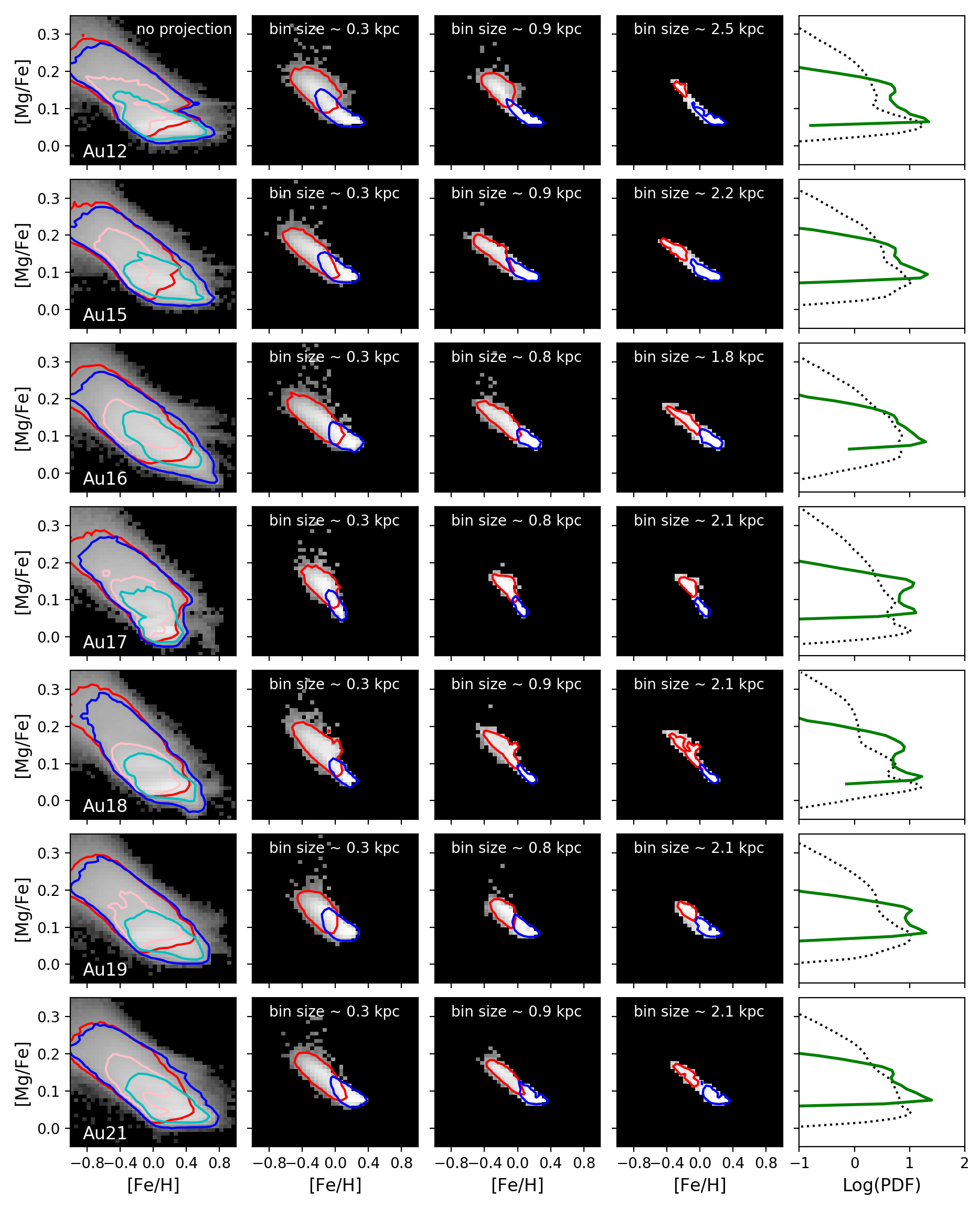}}
\caption{Same as Fig.~\ref{fig:bimod_4gal} for galaxies Au12, Au15 to Au19, and Au 21.
}
\label{fig:bimod_sample1}
\end{figure*}

\begin{figure*}
\centering
\resizebox{1.\textwidth}{!}
{\includegraphics[scale=1.]
{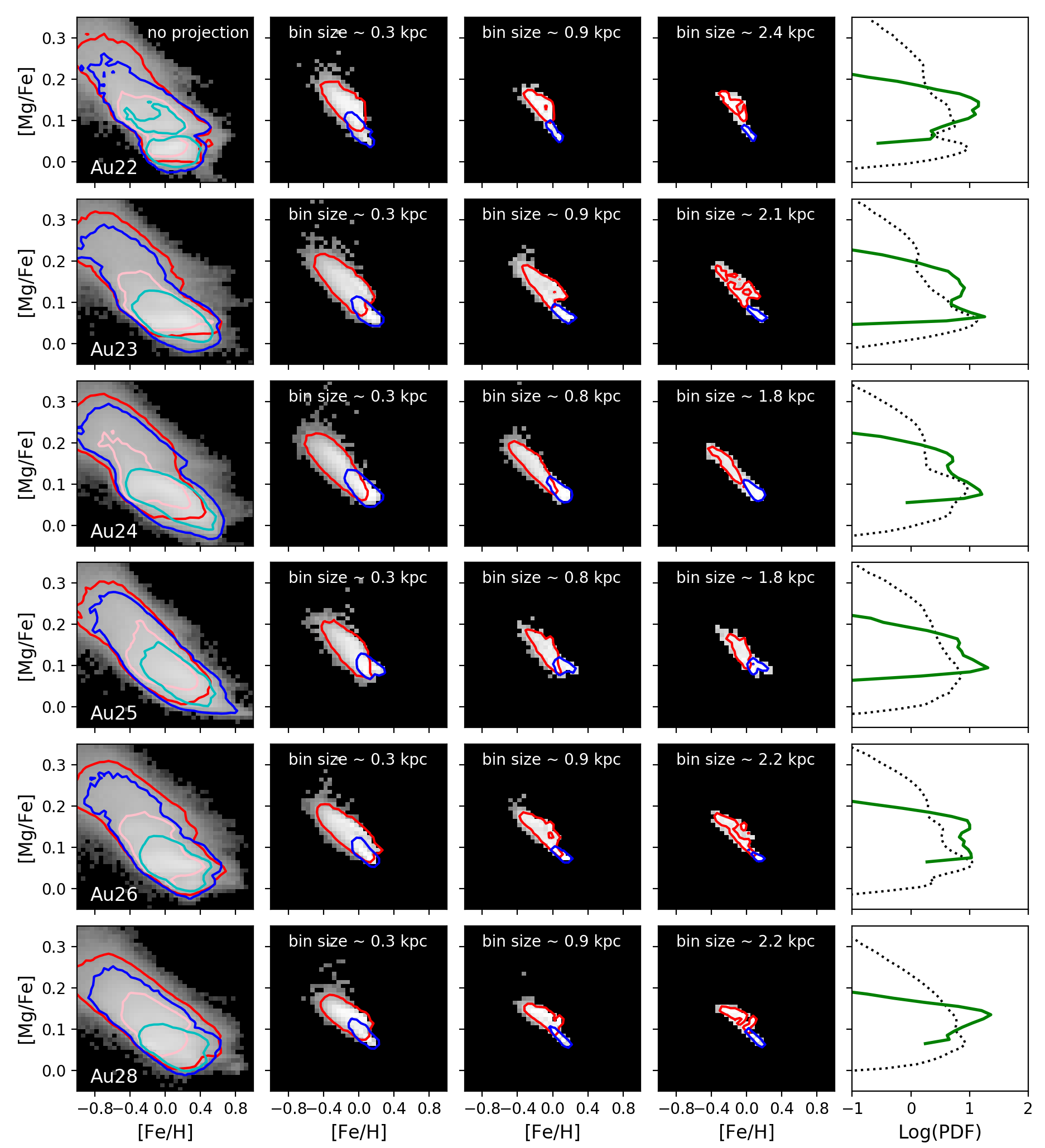}}
\caption{Same as Fig.~\ref{fig:bimod_4gal} for galaxies Au22 to Au26 and Au28. 
}
\label{fig:bimod_sample2}
\end{figure*}

We show similar plots to Fig.~\ref{fig:bimod_4gal}, for the remaining 20 galaxies of our sample, in Fig.~\ref{fig:bimod_sample0} to \ref{fig:bimod_sample2}. 
Based on these figures, we classified galaxies in Sect.~\ref{sub:class}. 
While this classification facilitates the analysis of the full sample, we note that each galaxy shows its own peculiarities. Moreover, galaxies classified in two different groups do not have necessarily very different distributions in the [Mg/Fe] - [Fe/H] plane. 
For instance, Au12, in Group 1 because the high-$\alpha$ component forms a small peak in the particle distribution, does not have a remarkably different particle distribution from Au15, in Group 3 since the high-$\alpha$ component is not clearly separated. 
Au5 can be classified in Group 1, as it displays clear bimodality in the Voronoi bin distribution and two distinct components in the particle distribution. However, one of these two components (the high-$\alpha$ one) has a hump shape (not a peak) and does not look clearly distinct. Therefore, Au5 has a particle distribution also similar to some galaxies in Group 3 (e.g., Au15). 

%

Au22 could be classified in Group 2 since the particle distribution has three components. However the Voronoi-bin distribution shows a roughly unimodal distribution with three small peaks at different [Mg/Fe]. The two higher-$\alpha$ peaks include stars located in the thick disk, while the lowest-$\alpha$ peak corresponds to the thin disk. 
Au28 shows a secondary small peak at lower [Mg/Fe], in the Voronoi-bin distribution, but not a clear bimodality, and is classified in Group 4. 
In Au14 (Fig.~\ref{fig:bimod_4gal}), the thin disk is made up of particles in two sequences, while in Au26 it is the thick disk to show two components with slightly different [Mg/Fe] in the distribution of individual particles. In Au10, Au17 and Au18, a central peak (at [Mg/Fe]$\sim 0.1$~dex) contributes to both particle distributions of the thick and thin disks. 

Au1 is a peculiar case and it has not been classified in any group. It shows three components in the particle distribution, two of them forming distinct sequences overlapping in [Fe/H]. 
These components probably correspond to three main formation phases of its disk. This galaxy had an intense merger history, almost 50\% of its stars were formed in satellites \citep{Pinna2023}, and it has a close massive companion which is still interacting with \citep{Grand2017}. 
The presence of two lower-$\alpha$ sequences points to two subsequent formation phases in which the last important merger played a key role (see the formation of a low-$\alpha$ thinner disk in Fig.~B.10 of \citealt{Pinna2023}). 
However, the different sequences and peaks in the particle distribution shrink to a unimodal narrow distribution, when transitioning from individual particles to mock IFS observations. 
Since in general thick and thin disks correspond to the two peaks of bimodal distributions (Sect.~\ref{sub:thickthin}), this supports that this galaxy does not have two clearly distinct morphological (thin and thick) disk components according to the analysis by \citet{Pinna2023}. 
Also, this galaxy shows that the presence of multiple sequences in the $\alpha$ - [Fe/H] plane does not necessarily lead to the formation of a double-disk morphological structure. 
This is the only case where bimodality is lost when stellar properties are projected and integrated into Voronoi bins. This indicates that all those [Mg/Fe] (and [Fe/H]) components are approximately evenly distributed in different regions (in particular, at different heights) of the galaxy, and overlapped in the line of sight when projected edge on. 
The intense merger history may have led to well-mixed properties diluting differences between regions (see also stellar-populations maps from \citealt{Pinna2023}).

\end{appendix}
\end{document}